\documentstyle[twocolumn,prl,aps]{revtex}
\begin{document}
\twocolumn[\hsize\textwidth\columnwidth\hsize\csname 
@twocolumnfalse\endcsname
\title{Attracting Manifold for a Viscous Topology Transition}
\author{Raymond E. Goldstein$^1$, Adriana I. Pesci$^2$, Michael J. Shelley$^3$}
\address{$^1$Department of Physics, Joseph Henry Laboratories,
Princeton University, Princeton, NJ  08544}
\address{$^2$Department of Physics, Middlesex College, 
Edison, NJ  08818}
\address{$^3$Courant Institute of Mathematical Sciences, New York University,
New York City, NY 10012}


\maketitle

\begin{abstract}

An analytical method is developed describing the approach to a
finite-time singularity associated with collapse of a narrow fluid layer 
in an unstable Hele-Shaw flow.  Under the separation of time scales near 
a bifurcation point, a long-wavelength mode entrains higher-frequency modes, 
as described by a version of Hill's equation.  In the slaved dynamics, 
the initial-value problem is solved explicitly, yielding the time and 
analytical structure of a singularity which is associated with the motion 
of zeroes in the complex plane.  This suggests a general mechanism of 
singularity formation in this system.

\end{abstract}
\vskip2pc]

One of the most fundamental questions underlying a broad class of
hydrodynamic phenomena is: How do smooth initial conditions evolve to
form finite-time singularities?  Historically, two main classes of
systems and phenomena have been investigated in this context
\cite{Crete}.  The first involves distributions of vorticity evolving
under Euler's equation \cite{Moore}.  The second, dating back
to classical work of Rayleigh \cite{Rayleigh},
concerns the motion of interfaces bounding masses of fluid undergoing
fission \cite{Stone,C92,GPS,Bertozzi,Eggers,Shi,Andy,Tanveer},  
or in other problems of pattern formation \cite{Shraiman}.
An issue on which considerable progress has been made is the 
relevance of self-similar solutions that may describe the region
asymptotically close to the topology transition.
However, there is little understanding of how such special solutions
emerge from the large scale dynamics.

We report here an approach to this {\it initial-value problem}
associated with a topological rearrangement of fluid interfaces.
Under the assumptions of lubrication theory, 
an approximation valid for long-wavelength deformations of thin layers,
the method is developed for the
Rayleigh-Taylor instability of two-phase Hele-Shaw flow \cite{GPS,Tryg}.   
This lubrication approximation has been the focus of a
considerable body of recent theoretical work on
thin film flows and the spreading of drops \cite{C92,GPS,Bertozzi,Eggers,Shi}.
Very similar, and even identical equations of motion describe 
phenomena as diverse as Marangoni convection \cite{Marangoni}, pattern formation 
in superconductors \cite{Chapman} and in biological systems \cite{Lewis}, and 
oxidation of semiconductor surfaces \cite{King}.

The method exploits the separation of time scales
occurring close to the first instability in a system of finite lateral
extent, where the spectrum of modes is discrete.  As in the study of
normal form expansions near convective or lasing instabilities
\cite{Cross_Hohenberg}, the slaving of the high-frequency modes
allows the derivation of a nonlinear
evolution equation for the amplitude of the first unstable mode.  It
also allows an analytic approximation of the singular contribution
from all other modes. 

{\it Dynamics and Separation of Time Scales}: We begin with the
equation of motion for the half-thickness $h(x,t)$ of a thin layer of
fluid in Hele-Shaw flow, bounded from above and below by mutually
immiscible fluids.  The equation of motion in rescaled form is
\cite{GPS}
\begin{equation}
h_t=-\partial_x\left(h\left[h_{xxx}+Bh_x\right]\right)~,
\label{lub_rt}
\end{equation}
where the Bond number $B=2g\Delta\rho L^2/\sigma$ is the single
dimensionless parameter characterizing the competition between the
surface tension $\sigma$ and buoyancy associated with density
jump $\Delta \rho$ in a gravitational field $g$, and where
$L$ is the lateral width of the Hele-Shaw cell.  
An unstable density stratification corresponds to $B>0$ \cite{GPS}.
The solution $h(x,t)$ can also be
interpreted as the height of a fluid layer lying at a wall, and
beneath another fluid.  Eq. \ref{lub_rt} is also relevant to recent 
experiments on the pinching of annular rings of fluid in the 
Hele-Shaw cell \cite{Andy}.  

As in related earlier models with $B=0$ \cite{C92,Bertozzi},
this PDE has a flux form $h_t+\bbox{\nabla}\cdot {\bf j}=0$ arising from 
incompressibility, with ${\bf j}=h{\bf U}$ the hallmark of lubrication theory.  
The characteristic
velocity ${\bf U}\sim -\bbox{\nabla} P$ arises from Darcy's law, 
and the pressure $P$ is
set by boundary conditions involving surface tension and gravity.  
In other contexts, the velocity has the more general form ${\bf U}\sim
-h^m\bbox{\nabla} P$, such as in the spreading of drops ($m=2$).  
Eq. \ref{lub_rt} with a different physical meaning to $B$ arises in 
the dynamics of the population density $h$ of feeding herbivores \cite{Lewis},
and also in the long-wavelength limit of the
homogenized model of Type-II superconductors \cite{Chapman}, with
$h$ the local density of vortices.

It is sufficient for our purposes to consider
Eq. \ref{lub_rt} in a system of length $2\pi$ with periodic boundary
conditions.  If $h(x,t)$ is even and periodic, then it also 
describes a flow between two rigid walls at which the interface 
has $90^{\circ}$ contact angle.
Linearizing about a flat interface $h=\bar h$ we obtain the growth rates
\begin{equation}
\nu(k)=\bar h\left(Bk^2-k^4\right) \ \ \ \ \ \ \ (k=0,1,2,\ldots)~.
\label{growth}
\end{equation}
The number $m$ of unstable modes scales as $\sqrt{B}$.
If $B<1$ all modes are linearly stable, whereas for $1<B<4$ the mode
$k=1$ is unstable, while those with $k\ge 2$ remain damped.  
Moreover,
if $B$ is tuned to be slightly larger than unity, say
$B=1+\epsilon$, then the first mode evolves on a time scale of
order $\epsilon^{-1}$, while the others rapidly equilibrate, thus
being {\it slaved} to the first.

{\it A Contracting Flow for $B=1$}:  Exactly at the bifurcation point
$B=1$ there is a {\it manifold} of steady-state solutions, $h_0(x)=\bar
h(1+a\cos x)$, for any $a < 1$.  This manifold is also an
attractor.  Let $h = h_0 + \delta \zeta_x$ ($\delta \ll 1$), with $\zeta$ of zero
mean.  The linearized evolution about $h_0$ is $\zeta_t = - h_0{\cal
L}_1\zeta_x$, where ${\cal L}_B=\partial_{xxx}+B\partial_x$, and
it preserves $\left<\zeta\right>=0$.  Consider then the norm
\[
{\cal F} = \int_0^{2\pi}\!\! dx {\zeta^2(x,t)\over 2 h_0(x)}
\ \ \ {\rm with}\  {\cal F}_t= -\sum_{k} k^2(k^2-1) \vert
{\hat\zeta}_k\vert^2 ~.
\]
One may verify the inequalities ${\cal F}_{t}
\ge {\cal F}_t(0)\exp(\pm H t) $, where $H=2\ {\rm min}_x\{h_0(x)\}>0$
\cite{GPS_long}. Since ${\cal
F}_t \le 0$, and further, ${\cal F}_t=0$ iff $\zeta$ is entirely in
the null space of ${\cal L}_1$ (and then $\zeta$ is a
steady state solution).  One inequality from above then
gives the bound
${\cal F}_t\ge{\cal F}_t(0)~e^{-Ht}$,
where ${\cal F}_t(0)\le 0$.  And so, ${\cal F}_t \rightarrow 0$ as
$t\rightarrow\infty$, which proves that the nullspace of ${\cal L}_1$
is attracting. 
Actually, $h_0=\bar h(1+a\cos kx)$ is a steady-state
for any integer $k$ when $B=k^2$, but is 
unstable to subharmonic perturbations if $k>1$.

{\it Motion along the Manifold for $B>1$}~:  As the Bond
number is increased slightly beyond unity, the first mode grows in time, 
but we expect that the amplitudes of
higher modes will remain small.  
More generally, for larger $B$, we expect a finite number $m$ of active modes 
including those that are linearly unstable.  A natural approach then
is to partition
$h$ into low ($p$) and high ($q$) modes.  Let ${\cal P}_m$ project
a periodic function onto its lower $m$ modes, and write 
\begin{equation}
h=p+q~,\ \ \ \ \ \ \left({\cal P}_m p = p~,~~~{\cal P}_m q = 0\right)~.
\label{low_high_modes}
\end{equation}
Substituting this decomposition into (\ref{lub_rt}), ignoring contributions 
of order $q^2$, invoking slaving of higher modes, ($\partial_t q\simeq 0$),
and integrating twice with respect to $x$, we obtain
\begin{equation}
p q_{xx} - p_x q_x + (p_{xx}+Bp)q =-{\widetilde{\widetilde p}}_t
-{\widetilde J_p}+C~,
\label{Ince}
\end{equation}
where $J_p=p{\cal L}_B p$ is the {\it flux} associated with the lower
modes, $C$ is an integration constant, and $\widetilde f=\int^x\! dx'
f(x')$.  Since $p$ is periodic in $x$, we find rather
remarkably the computation of $q$
reduced to the solution of an inhomogeneous Hill's equation
\cite{Magnus}.  Coupling to the partition constraints, ${\cal P}_m p_t
= p_t$ and ${\cal P}_m q = 0$, gives a complete set of equations to
determine $q$ and $p_t$.

The {\it slaving approximation} (\ref{low_high_modes}) and (\ref{Ince}) is
particularly easy to analyze in the limit $B-1\equiv \epsilon \to 0$, 
for which there is only one active mode, thus 
$p=\bar h\left(1+a(t)\cos x\right)$,
and we rescale (\ref{Ince}) with 
\begin{equation}
B=1+\epsilon~; \ \ \ \tau=\epsilon \bar h t~; \ \ \ q=\epsilon Q~.
\label{rescalings}
\end{equation} 
At ${\cal O}(\epsilon)$ we obtain an inhomogeneous Ince's equation \cite{Magnus},
\begin{eqnarray}
\left(1+a\cos x\right)Q_{xx}&+&a\sin x ~Q_x+Q\nonumber \\
&=&\!a_{\tau}\cos x-{1\over 2}\left(1+a\cos x\right)^2+C~,
\label{Ince_1}
\end{eqnarray}
where $C=(1+a^2/2)/2$ by the orthogonality of $p$ and
$q$.  

{\it Solvability and the Amplitude Equation}:  
The solution of Eq. \ref{Ince_1}, found by variation of parameters, 
contains a secular term proportional to $x\sin x$.  Its removal 
is the solvability
condition that determines $a_{\tau}$, and yields the nonlinear
equation of motion
\begin{equation}
a_{\tau}={a\over 2}\left(1+\sqrt{1-a^2}\right)~.
\label{a_eqnofmotion}
\end{equation}
For $a\ll 1$ this yields the exponential growth
$a_{\tau}=a+\cdots$ of the linear stability result
(\ref{growth}), but this behavior crosses over to a much different
form in the nonlinear regime near pinching.  Defining the 
function
\begin{equation}
f(a)= {1-\sqrt{1-a^2}\over a^2}
-\log\left({1-\sqrt{1-a^2}\over a}\right)~,
\label{f_define}
\end{equation}
with $a_0=a(\tau=0)$, we find $f(a_0)-f(a)=\tau$ by direct integration
of (\ref{a_eqnofmotion}).  The singularity (or ``pinch") time $\tau_p$
occurs when $a\nearrow 1$, so $f(a_0)-1=\tau_p$, and thus 
\begin{equation}
t_p= {f(a_0)-1\over \bar h(B-1)}~.
\label{a_of_t}
\end{equation}
For $a_0\ll 1$, $t_p\sim \log(2/a_0)/\bar h(B-1)$, again consistent
with the exponential growth of the linear instability.
Near the touchdown, $a(t)=1-\left(t_p-t\right)/2 +\cdots~$.  Figure
\ref{Fig1} shows excellent agreement between these asymptotic results and
numerical studies of the lubrication PDE (\ref{lub_rt}) for
the pinch times $t_p(a_0)$, and for the minimum height $h_{\rm min}=\bar
h(1-a(t))$.

The correction $Q$ is found to be
\begin{eqnarray}
Q(x)&=&\lambda_+\Biggl\{\sqrt{1-a^2}\sin x
\tan^{-1}\left({\lambda_-\sin x\over 1-\lambda_-\cos x}\right)\nonumber \\
&&\qquad-{1\over 2}\left(a+\cos x\right)\ln\left(1-2\lambda_-\cos x+
\lambda_-^2\right)\nonumber \\
&&\qquad-a\left({3\over 4}\lambda_-\cos x-{1\over 2}\right)\Biggr\}~,
\label{h_great}
\end{eqnarray}
where $\lambda_{\pm}$ are the two real {\it zeroes} of the quadratic
$a\lambda^2+2\lambda+a=0$, for which $\lambda_{+}\lambda_{-}=1$, and
$\lambda_{-}\le 1$.  As $a\nearrow 1$, $\lambda_-\to -1$, and thus 
within this analysis the interface curvature,
through $Q''(x)$, develops a logarithmic singularity.  This divergence can 
also be interpreted as the
collision on the real axis of two singularities, located at $\pi \pm i
\ln \left| \lambda_{-} \right |$ in the complex $x$-plane.

{\it A Spectral Cascade}:  The finite-time singularity
can be understood by considering the spectrum of $Q$, writing
$Q=\sum_{k=2}^{\infty} Q_k\cos kx $. The recursion
relation for the $Q_k$'s obtained from 
Eq. \ref{Ince_1} has the {\it exact} solution (for $k\ge 3$)
\begin{equation}
Q_k=C_+ {\lambda_+^k\over k^3-k} + C_-{\lambda_-^k\over k^3-k}~,
\label{recursion_soln}
\end{equation}
where $C_{\pm}$ are constants determined by the inhomogeneous terms.
Since $\vert \lambda_+\vert>1$ for $a<1$, the first term in
(\ref{recursion_soln}) is the secular term eliminated by the
solvability condition, and thus the power-law spectrum of $Q$
is cut off by an exponential factor which tends to unity
as $a\nearrow 1$.
Full simulations show very good agreement, to very near the singularity time,
between the form of the
correction function (and its spectrum) with the asymptotic 
result (\ref{h_great}).  
Figure \ref{Fig2}a shows a comparison between the two in real space,
and Fig. \ref{Fig2}b shows the strong agreement between pointwise estimates
$-\ln\vert\lambda\vert\simeq
\ln(Q_k/Q_{k+1})$ for four wavenumbers $k\gg 1$,  illustrating
the collapse of the analyticity strip width in accord with
Eq. \ref{a_eqnofmotion}. 
At extremely small values of $h_{\rm min}$, the
slaving assumptions should break down, and terms such as $q_t$ cannot be
neglected.  Indeed, A. Bertozzi has noted that the 
ultimately negative divergence of
$Q_{xx}$, while only logarithmic, is inconsistent with the existence of a
single touchdown \cite{Bertozzi_note}.  Her numerical studies  
following $h_{\rm min}$ down to ${\cal O}(10^{-30})$ suggest instead a saturation 
of the curvature.

{\it Bifurcation of the Singularities}:  Finally, we consider 
values of the Bond number well beyond the bifurcation
point $B=1$.  Using an initial condition $h=\bar h\left(1+
a\cos x \right)$, with $a=0.01$, Figure 3 shows how the single {\it
symmetric} touchdown at $x=\pi$ seen for $B\gtrsim 1$ bifurcates for
$B\gtrsim 1.55$ into two {\it asymmetric} touchdowns.  This behavior
reflects the general increase in the number of active modes with increasing $B$.
While the asymptotics
developed for $B=1+\epsilon$ are not quantitatively valid for
$\epsilon ={\cal O}(1)$, the slaving hypothesis with an increased number of
modes leads to an appealing picture of how the singularities are generated in this
regime.

The ansatz $p=\bar h\left(1+a(t)\cos x+b(t)\cos 2x \right)$
is the simplest allowing for two singularities, and leads to a spectrum
of the form \cite{GPS_long} 
\begin{equation}
Q_k\sim\sum_{\nu=1}^4 C_{\nu}{\lambda_{\nu}^k\over k^3} \ \ \ \ \ \ \ (k\gg1)~,
\label{two_mode_Q}
\end{equation}
where $\{\lambda_{\nu}\}$ are the zeroes of the quartic 
$b\lambda^4+a\lambda^3+2\lambda^2+a\lambda+b=0$, defined by
two independent quantities $\lambda_{1,2}$ as
$\lambda_1,\lambda_1^{-1},\lambda_2,\lambda_2^{-1}$.  Except when any
$\vert \lambda_{1,2}\vert=1$, two of the four zeroes lie within the
unit circle in the complex $\lambda$ plane, the other two lie outside.
Elimination of the secular solutions associated with the latter
two yields $a_t(t)$ and $b_t(t)$.  As $a$ and $b$ evolve in time,
the remaining zeroes move toward the unit circle.  In the first
quadrant of the $(a,b)$
parameter space the line $b=a-1$, for ($1\le a \le 4/3$) defines those
pinching configurations with a single touchdown (at $x=\pi$), while
the curve $a=[8b(1-b)]^{1/2}$ for $a>4/3$ is the locus of
configurations with two touchdowns.  In the former case,
$\lambda_{1,2}$ are real (as are $C_{1,2}$), and only one zero reaches the
unit circle at the pinch time, thus producing only one singularity.
Beyond point $a=4/3,b=1/3$, $\lambda_1$ and $\lambda_2$ are complex
conjugates of each other (with $C_1=C^*_2=R~e^{i\psi}$) and reach
the unit circle simultaneously, producing two singularities.  

It follows from this analysis that any Ansatz for the active modes
contained in $p$ generates a set of zeroes in the complex plane.  
Some of these singularities will
move toward, although not all reach the unit circle as the pinch time is approached.
Simulations of the full problem agree well with this picture. 
The numerical studies indicate that the nature of the singularities
depends on the symmetry of the touchdowns;
we observe a jump discontinuity in $h_{xx}$ for asymmetric pinching, 
rather than the divergence seen with a symmetric pinch.
This is associated with a rotation of the phase $\psi$ to $\pi/2$.  
We do not yet understand whether this behavior
may be captured within the slaving approximation.  

This analysis thus merges two previously separate concepts in dynamical
systems described by PDEs. First is the coupling of slaved small scales to low-mode
dynamics that recalls the reduction of a dissipative PDE to an inertial 
manifold \cite{inertial}. Second is the motion of  
zeroes in the complex plane as in the reduction of certain PDEs to
``pole dynamics" \cite{Shraiman,Kruskal}.

We thank A. Bertozzi, R. Kohn, and M. Pugh for discussions.  This work
was supported in part by NSF
PFF grant DMR93-50227 and the A.P. Sloan Foundation (REG), by NSF 
PYI grant DMS-9396403, grant
DMS-9404554, DOE grant DE-FG02-88ER25053, and by the Exxon Educational
Foundation (MJS).

\begin{figure}
\caption{Comparison between numerical
solution of lubrication equation (\protect{\ref{lub_rt}}) and asymptotic 
analysis for $B\to 1$.
(a) Singularity time versus initial amplitude $a_0$ from numerical solution of
Eq. \protect{\ref{lub_rt}} (solid circles) for $B=1.05$, and from
asymptotics in Eq. \protect{\ref{a_of_t}}.
(b) Minimum interface height as a function of time.
Solid lines
are the results of Eq. \protect{\ref{a_eqnofmotion}} with $a_0=0.01, 0.05, 0.30$
from top to bottom of figure, all with $B=1.05$; solid circles 
show numerical results for those same initial conditions.}
\label{Fig1}
\end{figure}

\begin{figure}
\caption{(a) The function $Q(x)$ in Eq. \protect{\ref{h_great}} 
obtained from the asymptotic analysis (solid), compared with
numerical solution of the full PDEs for $B=1.05$ (dots).
Results are for four times ranging from close to the initial condition
to near the singularity time. (b) Collapse of the analyticity strip width
as a function of time, from numerical studies.
For the largest two values of $k$ shown,
deviations from the common curve arise from these amplitudes lying 
initially beneath machine precision.}
\label{Fig2}
\end{figure}

\begin{figure}
\caption{Bifurcation diagram showing singularity locations versus Bond number.
Insets (a) and (b) show interface evolution at $B=1.25$ and $B=2.0$.}
\label{Fig4}
\end{figure}

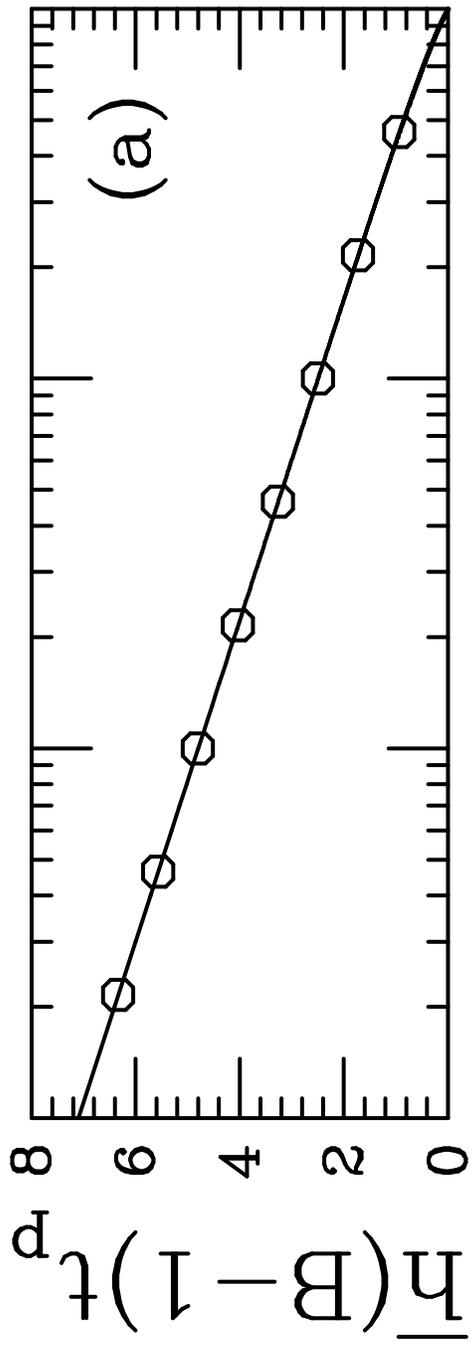
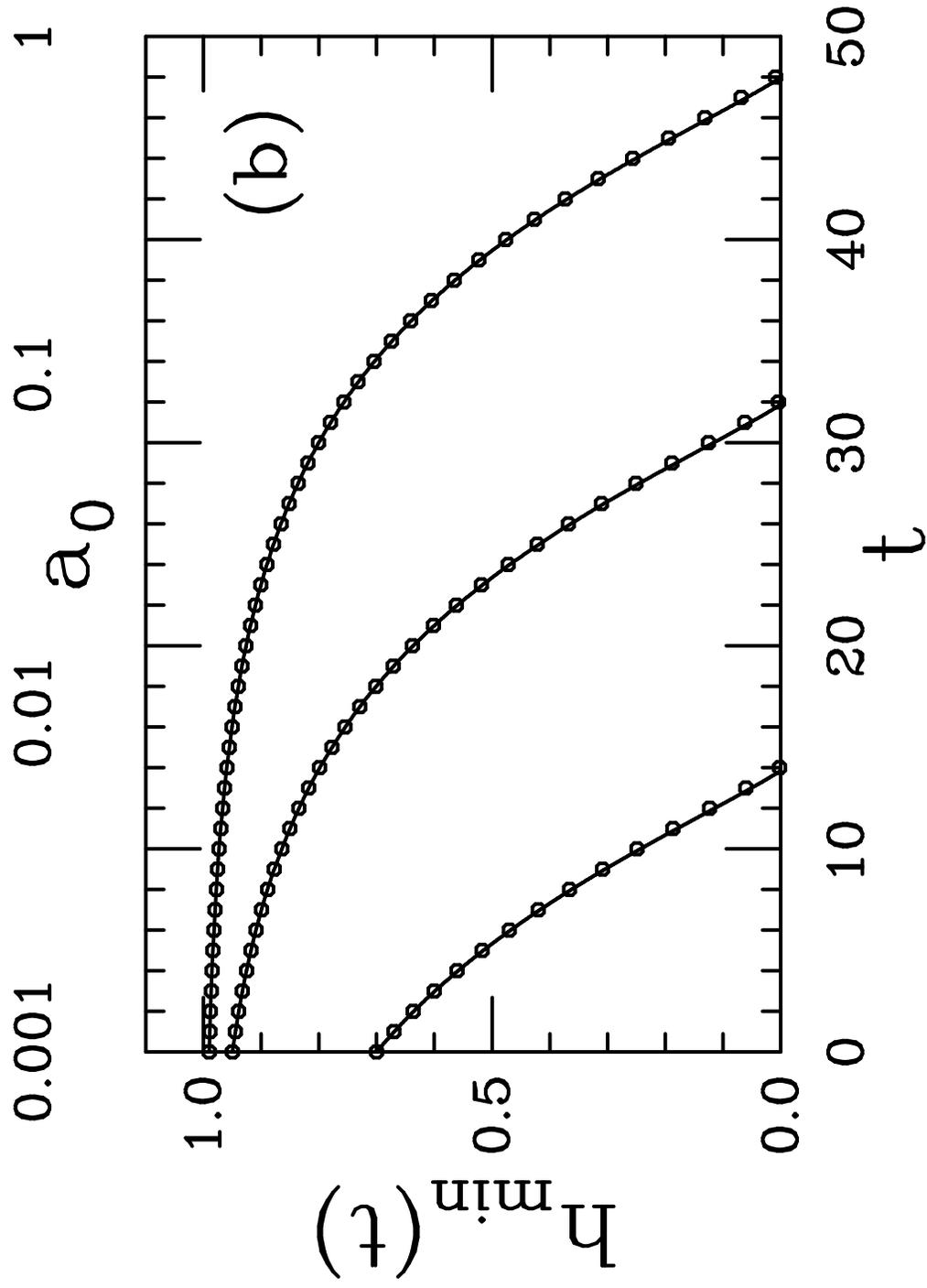
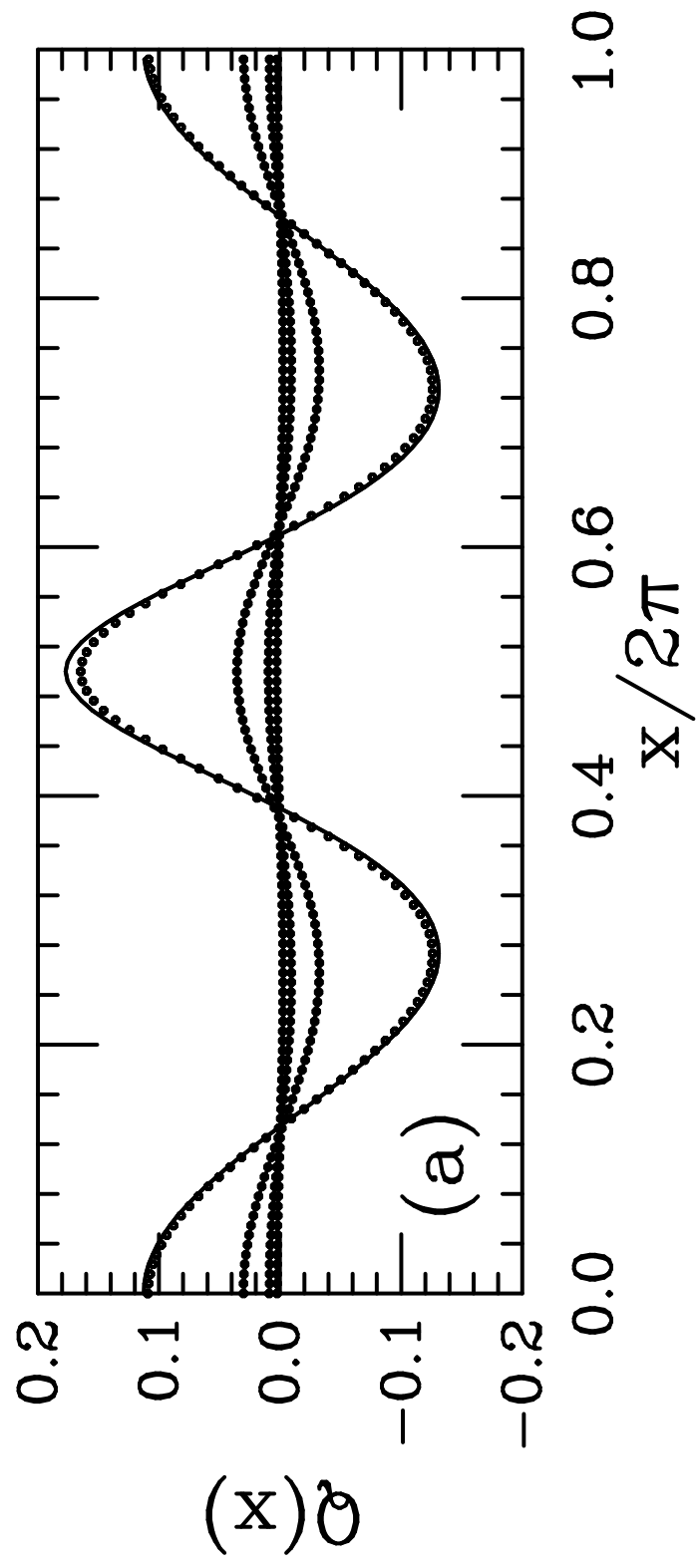
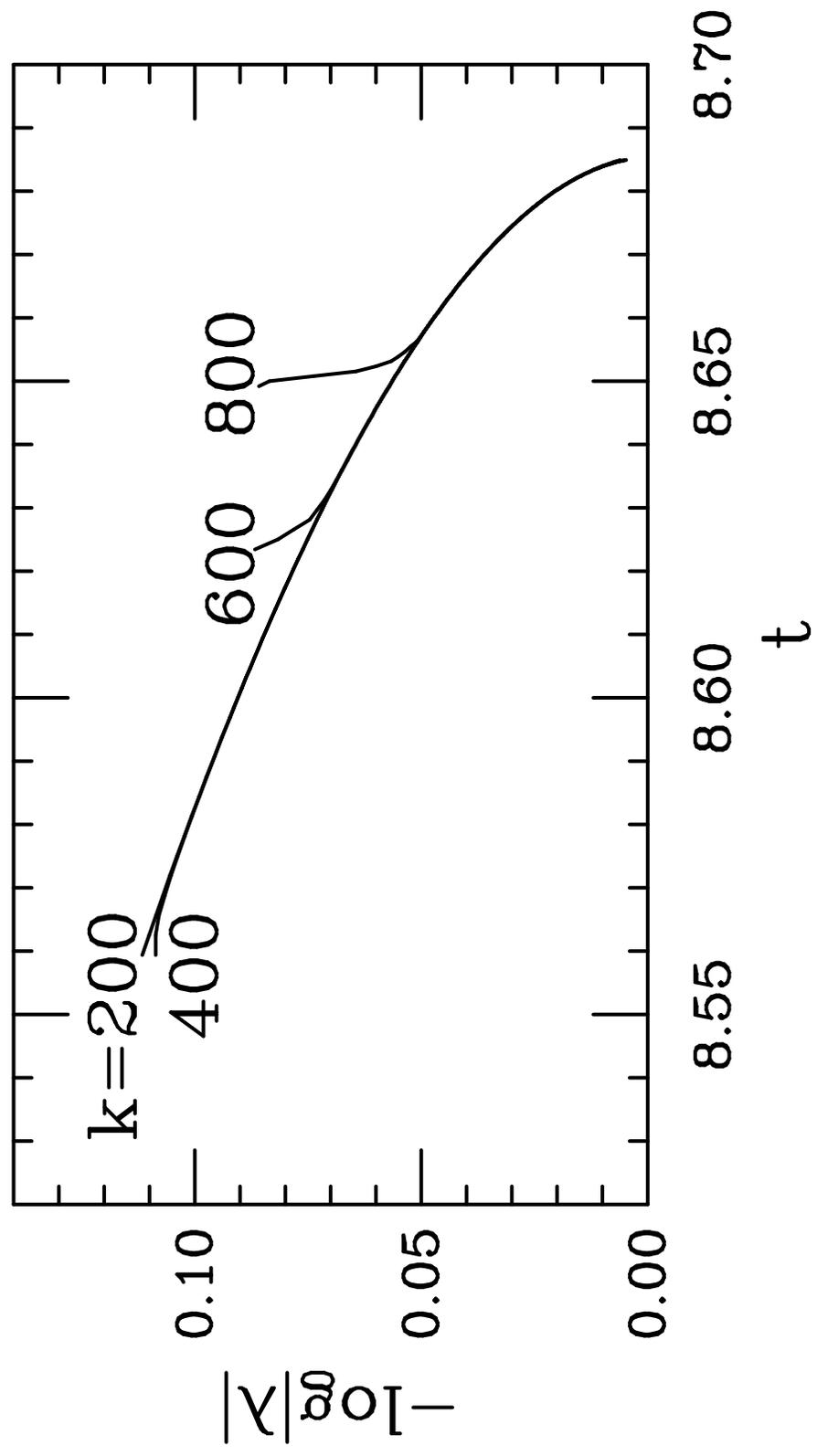
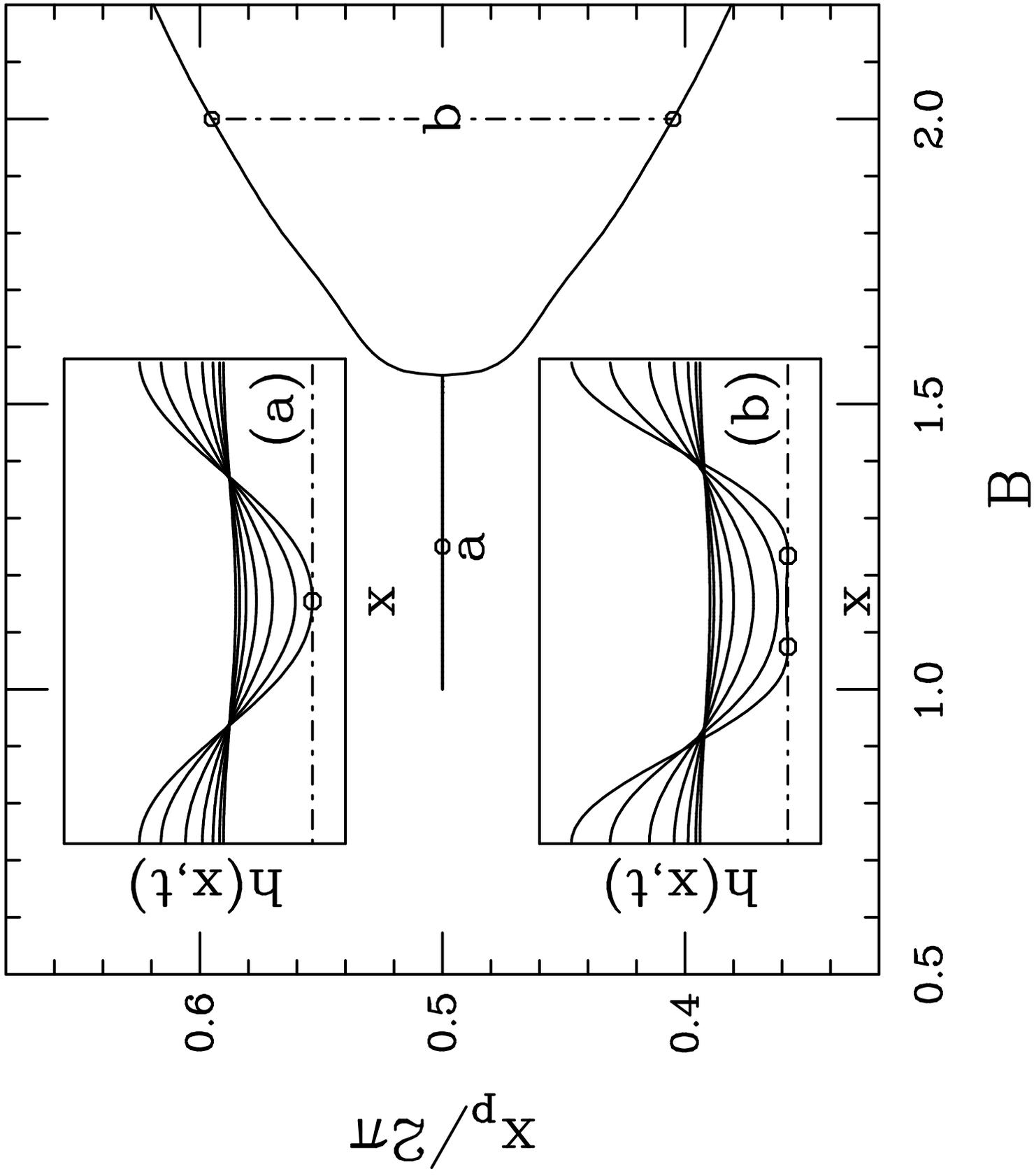
\end{document}